\documentstyle[12pt,cite,epsfig]{article}
\newcommand{\sect}[1]{\setcounter{equation}{0}\section{#1}}

\newcommand{\be}{\begin{equation}}
\newcommand{\ee}{\end{equation}}
\newcommand{\bea}{\begin{eqnarray}}
\newcommand{\eea}{\end{eqnarray}}

\newcommand{\p}{\partial}

\def\IB{\relax\hbox{$\inbar\kern-.3em{\rm B}$}}
\def\IC{\relax\hbox{$\inbar\kern-.3em{\rm C}$}}
\def\ID{\relax\hbox{$\inbar\kern-.3em{\rm D}$}}
\def\IE{\relax\hbox{$\inbar\kern-.3em{\rm E}$}}
\def\IF{\relax\hbox{$\inbar\kern-.3em{\rm F}$}}
\def\IG{\relax\hbox{$\inbar\kern-.3em{\rm G}$}}
\def\IGa{\relax\hbox{${\rm I}\kern-.18em\Gamma$}}
\def\IH{\relax{\rm I\kern-.18em H}}
\def\IK{\relax{\rm I\kern-.18em K}}
\def\IL{\relax{\rm I\kern-.18em L}}
\def\IP{\relax{\rm I\kern-.18em P}}
\def\IR{\relax{\rm I\kern-.18em R}}
\def\IC{\relax{\rm I\kern-.18em C}}
\def\IZ{\relax{\rm Z\kern-.5em Z}}

\begin{document}
\topmargin 0pt
\oddsidemargin 0mm

\renewcommand{\thefootnote}{\fnsymbol{footnote}}
\begin{titlepage}

\begin{flushright}
IP/BBSR/2003-36\\
ROM2F/2003/30\\
hep-th/0310219\\
\end{flushright}

\begin{center}
{\Large \bf  More D-branes in plane wave spacetime}\\
\vskip .5in
{\bf Rashmi R. Nayak$^a$\footnote{e-mail: rashmi@iopb.res.in} and
Kamal L. Panigrahi $^{a, b}$ \footnote{$^b$ Present address 
(I.N.F.N. Fellow)}
\footnote{e-mail: kamal@iopb.res.in,~ Kamal.Panigrahi@roma2.infn.it}}\\

\vskip .5in
{\em $^a$ Institute of Physics,\\
Bhubaneswar 751 005, INDIA\\
\vskip .5in
$^b$ Dipartimento di Fisica,\\ 
Universita' di Roma ``Tor Vergata",\\
I.N.F.N.-Sezione di Roma ``Tor Vergata",\\
Via della Ricerca Scientifica, 1\\
00133  Roma  ITALY}
\end{center}
\vspace{0.1in}
\begin{center}
{\bf ABSTRACT}
\end{center}

\vskip .5in
\noindent
We present classical solutions of $Dp$-branes $(p\ge 5)$ in plane wave 
spacetime with nonconstant R-R 3-form flux. We also show the existence of
a system of $D3$-branes in this background. 
We further analyze the supersymmetric properties of these
branes by solving type II Killing spinor equations explicitly.

\vfill

\end{titlepage}

\section{Introduction}
Study of string theory in plane wave background with flux has been 
the topic of intense discussion in recent past. It is known for quite
sometime that pp-wave spacetime provides exact string theory
backgrounds. These backgrounds are exactly solvable 
in lightcone gauge. Many of them are obtained in
the Penrose limit (pp-wave limit) of $AdS_p \times S^q$ 
type of geometry and in some cases are maximally supersymmetric
\cite{penrose,blau}. Strings in pp-wave background are also
investigated to establish the duality between the supergravity modes
and the gauge theory operators in the large R-sector of the gauge
theory\cite{malda}. 

PP-wave background with nonconstant Ramond-Ramond (R-R) flux 
\cite{maoz,russo,hikida,kim,bonelli}
gives an interesting class of supersymmetric pp-wave solutions
in type IIB supergravity. The worldsheet theory corresponding to pp-waves
with nonconstant R-R $F_5$ flux is described by nonlinear sigma model
which is supersymmetric and 
one can have linearly realized `supernumerary' supersymmetries in 
these backgrounds\cite{pope}.  
PP-wave backgrounds supported by nonconstant R-R $F_3$ fields,
do not have, in contrast to their $F_5$ counter part, supernumerary 
supersymmetries. These backgrounds provide, in general, examples of
nonsupersymmetric sigma models\cite{russo} unless there exists
some target space isometry and corresponding Killing vector potential
terms, which ensure the worldsheet supersymmetry\cite{kim}. 
The bosonic string action of a general class of pp-wave background supported
by nonconstant R-R $F_5$ flux, in light cone gauge, can be read off from
the metric. The nonlinear sigma models have eight dimensional special
holonomy manifold target space. The nonvanishing R-R fields gives, in 
particular, fermionic mass terms in the worldsheet action.
Classical solutions of $D$-branes in pp-wave background with
constant NS-NS and R-R flux are already discussed in the literature
\cite{kumar,sken,bain,alis,zamak,kamal,rashmi,alday}. $Dp$-branes 
from worldsheet point of view are constructed in \cite{dabh}.
Supersymmetric properties of $D$-branes and their bound states 
have also been analyzed both from supergravity and from worldsheet 
point of view.

$D$-branes and their bound states play an important role in understanding
various nonperturbative and duality aspects of string theory and gauge 
theories. The configurations of branes oriented by certain $SU(N)$ 
angle are known to be supersymmetric objects
\cite{douglas,gauntlett,cvetic,myers,hambli,leigh,townsend,ohta}. 
They have also been useful in
understanding the physics of black holes and gauge theories.
So it is worth examining various classical solutions $D$-brane
in plane wave spacetime as they also represent black holes in these
backgrounds. 
The pp-wave spacetime with nonconstant five form flux has 
the interpretation of soliton solutions in two- dimensional 
sigma models as emphasized by Maldacena-Maoz\cite{maoz}.
(For a recent related work see \cite{alin}). So a natural extension
would be to consider $D$-branes in these and in more general background
to find out the interacting nonlinear sigma models on the worldsheet
in the presence of $D$-branes. So it is desirable to 
study various supergravity solutions of $D$-branes in order to have 
the spacetime realization of these objects and to study their
supersymmetry properties as well.        

In earlier work, we found some classical solutions of $D$-branes along
with the supersymmetry in pp-wave spacetime with nonconstant 
NS-NS flux \cite{kamal1}. Intersecting $D$-branes in supergravities have also
been discussed in \cite{liu,ohta1}.
The possible black branes and the horizons have been discussed in the 
nonextremal deformations of $D$-branes in these backgrounds.
So it is interesting to find out more $D$-brane solutions in plane wave 
spacetime with flux and to discuss the possibility of horizons 
in this framework. In this paper, we continue the search for
supergravity brane solutions in plane wave spacetime with
nonconstant R-R $F_3$ flux. First we present the classical solutions of 
$Dp$-branes $(p\ge 5)$ in plane wave spacetime with nonconstant R-R $F_3$ 
flux. Next, we find classical solution of
a system of $D3$-branes oriented at an angle $\alpha,~(\alpha \in SU(2))$ with
respect to each other in this background. In the $D5$-brane case 
all the worldvolume coordinates of the brane lie along the
pp-wave directions and the transverse directions are flat. 
On the otherhand, for the $D3$-brane
system only lightcone directions are along the brane, whereas the other
pp-wave directions are along the transverse space. We would like to 
point out that the $D$-branes found in this paper are examples of
localized $D$-branes in plane wave spacetime with flux. We would also
like to point out that all the $D$-branes presented here are 
{\it longitudinal branes} as explained in \cite{bain}. The rest of the
paper in organized as follows. In section-2, we present classical solutions
of $D$-branes in pp-wave background with nonconstant R-R flux.
Section-3 is devoted to the supersymmetry analysis of brane solutions 
presented in section-2. We conclude in section-4 with some 
discussions.
 
\sect{Supergravity Solutions}

We start by writing down the supergravity solution of a system of
$D5$-branes in the pp-wave background with non-constant R-R 3-form flux.
The metric, dilaton and field strengths of such a configuration is given by:

\bea
ds^2&=&f^{-{1\over 2}}_5\left(2dx^+dx^- + K(x_i) (dx^+)^2+\sum^4_{i=1}
(dx_i)^2\right)
+f^{{1\over 2}}_5\left(dr^2+r^2d\Omega_3^2\right), \cr
& \cr
F &=& \p_1 b_2(x_i) ~dx^+ \wedge dx^1 \wedge dx^2 + \p_3 b_4(x_i) ~dx^+ \wedge 
dx^3 \wedge dx^4, \cr
& \cr
e^{2\phi} &=& f^{-1}_5,~~~~~~~F_{abc}=\epsilon_{abcd}\partial_{d}f_5, \cr 
& \cr
f_5 &=&1+{Ng_sl_s^2\over r^2},
\label{D5}
\eea
with $\Box K(x_i) + (\p_i b_j)^2 =0$ and $\Box b(x_i) = 0$. $f_5$ denotes
the harmonic function that satisfies Green function equation in 
the transverse 4-space. We have
checked that the solution presented above satisfies all type IIB field
equations. Other $Dp$-brane $(p \ge 6)$ solutions can be obtained by
applying $T$-duality along $x^5,...,x^8$ directions. For example:
the $D6$-brane solutions, by applying $T$-duality along $x^5$ (say), 
is given by:
\bea
ds^2&=&{f_6}^{-{1\over 2}}\left(2dx^+dx^- + K(x_i) (dx^+)^2+\sum^4_{i =1}
{(dx^i)}^2 + {(dx^5)}^2\right)
+{f_6}^{1\over 2}\left(dr^2+r^2d\Omega_2^2\right), \cr
& \cr
F &=& \p_1 b_2(x_i) ~dx^+ \wedge dx^1 \wedge dx^2 \wedge dx^5
+ \p_3 b_4(x_i) ~dx^+ \wedge dx^3 \wedge dx^4 \wedge dx^5, \cr
& \cr
e^{2\phi} &=& f^{-{3\over 2}}_6,
~~~~~~~F_{ab}=\epsilon_{abc}\partial_{c}f_6, \cr 
& \cr
f_6 &=&1+{Ng_sl_s\over r}.
\label{D6}
\eea
Where $f_6$ is the harmonic function that satisfies 
Green function equation in the transverse 3-space. 
Similarly, one can continue the above exercise for finding out 
supergravity solutions of the higher branes like $D7$ as well.
Bound states of $D$-branes can also be constructed by applying 
$T$-duality in the `delocalized' $D$-brane solutions as explained in
\cite{myers1,costa}. For example a $D5-D7$ bound state can be obtained from
a $D6$ solution and so on. We would like to point out that the solutions
presented here are the generalization of the $D$-brane solutions
found out in\cite{kumar}. However, the crucial difference
lies in the realization of supersymmetry, which will be discussed
in the next section.    

Now we present the classical solutions of a system of $D3$-branes oriented
at an $SU(2)$ angle with respect to each other in pp-wave background
with nonconstant R-R 3-form flux. First, we present the supergravity 
solution of a single $D3$-brane oriented at an angle $\alpha\in
SU(2)$ with respect to the reference axis. To start with, 
the $D3$-brane is lying along $x^+, x^-, x^6$ and $x^8$ directions. 
By applying a rotation between $(x^5-x^6)$ and $(x^7-x^8)$-planes
following \cite{myers},  with rotation angles $(\alpha_1, \alpha_2) =
(0, \alpha)$, we get a configuration where the original 
$D3$-brane is tilted by an angle $\alpha$. In stead 
of going more into the constructional details, below we write down the 
classical solution of a single $D3$-brane rotated by an angle $\alpha$: 
\begin{eqnarray}
ds^2 &=&\sqrt{1+X_1}\Bigg\{{1 \over 1+X_1}\bigg(
2{d x^+}{d x^-} + K(x_i){(d x^+)}^2 \cr
& \cr
&+& [1 + X_1 \cos^2\alpha][(d x^5)^2 + (d x^7)^2] 
+ [1 + X_1 \sin^2\alpha][(d x^6)^2 + (d x^8)^2] \cr
& \cr
&+& 2X_1 \sin\alpha \cos\alpha (d x^7 d x^8 -d x^5 d x^6) \bigg)
+ \sum_{i=1}^4 (d x^i)^2\Bigg\} \cr
& \cr
F &=& \p_1 b_2(x_i) ~dx^+ \wedge dx^1\wedge dx^2 + \p_3 b_4(x_i) ~dx^+ \wedge
dx^3 \wedge dx^4, \cr
& \cr
F^{(5)}_{+-68i} &=& - {\partial_i X_1\over (1 + X_1)^2}\cos^2\alpha,~~~~  
F^{(5)}_{+-67i} =  {\partial_i X_1\over (1 + X_1)^2}
\cos\alpha \sin\alpha, \cr
& \cr
F^{(5)}_{+-57i} &=& {\partial_i X_1\over (1 + X_1)^2}\sin^2\alpha,~~~~
F^{(5)}_{+-58i} = - {\partial_i X_1\over (1 + X_1)^2}
\cos\alpha \sin\alpha, \cr  
& \cr
e^{2 \phi} &=& 1.
\label{d3-1}
\end{eqnarray}
and $X_1$ is given by
\begin{equation}
X_1(\vec r) = { 1\over 2}
\bigg( {\ell_1 \over \vert \vec r - \vec r_1\vert}\bigg)^2.\label{d3-1a}
\end{equation}
Where $r$ is the radius vector in the transverse space, defined by
$r^2 = \sum^4_{i=1} (x^i)^2$, $r_1$ is the location of $D3$-brane and
$X_1$ is the Harmonic function in the transverse space. 
One can easily check that the above ansatz solve type IIB field
equations, with  $\Box K(x_i) = - (\p_i b_j)^2$ and $\Box b(x_i) = 0$.

Next, we present the supergravity solution of a system of two
$D3$-branes oriented at an angle $\alpha$ with respect to each other.
In this case, to start with two $D3$-branes are parallel to 
each other and are lying along $x^+, x^-, x^6$, $x^8$ directions. By
applying an $SU(2)$ rotation as described earlier, the second 
brane rotated by an angle $\alpha$, now lies along $x^+, x^-, x^5$ and
$x^7$ directions. The metric, dilaton and the field strengths of such
a system is given by: 
\begin{eqnarray}
ds^2 &=& \sqrt{1+X}\Bigg\{{1 \over 1+X}\bigg(
2{d x^+}{d x^-} + K(x_i){(d x^+)^2} \cr
& \cr 
&+& (1+X_2)\big[(d x^5)^2 + (d x^7)^2\big] + (d  x^6)^2 + (d x^8)^2 \cr
& \cr
&+& X_1\left[( \cos \alpha d x^5 - \sin \alpha d x^6)^2
+ ( \cos \alpha d x^7 + \sin \alpha d x^8)^2\right]\bigg)
+ \sum_{i=1}^4 (d x^i)^2\Bigg\},\cr
& \cr
F &=& \p_1 b_2(x_i) ~dx^1 \wedge dx^2 + \p_3 b_4(x_i) ~dx^3 \wedge dx^4, \cr
& \cr
F^{(5)}_{+-68i} &=& {\partial_i\Big\{{{X_2 + X_1 \cos^2\alpha +
X_1 X_2 \sin^2\alpha} \over {(1 + X)}}\Big\}},\cr
& \cr
F^{(5)}_{+-58i} &=& - F^{(5)}_{+-67i} = {\partial_i\Big\{{{X_1 \cos\alpha 
\sin\alpha} \over {(1 + X)}}\Big\}},\cr
& \cr
F^{(5)}_{+-57i} &=& - {\partial_i\Big\{{{(X_1+ X_1 X_2) \sin^2\alpha}
 \over {(1 + X)}}\Big\}},~~~~e^{2 \phi} = 1 .
\label{d3-2}
\end{eqnarray}
and $X$ is the Harmonic function in the transverse space which is given by
\begin{equation}
X =\, X_1 + X_2 + X_1 X_2 \sin^2 \alpha,\label{d3-2a}
\end{equation}
where as defined earlier, $X_{1,2} = { 1\over 2}
\bigg( {\ell_{1,2} \over \vert \vec r - \vec r_{1,2}\vert}\bigg)^2$.
Once again we have checked that the above solution solve type IIB 
field equations, with $\Box K(x_i) = - (\p_i b_j)^2$ and $\Box b(x_i) =
0$. More $D$-brane bound states can be obtained by applying $T$-duality
transformation along $x^5,...,x^8$ directions. We would like to
point out that the $D$-brane solutions presented here are the
generalizations of the solutions presented in \cite{myers,rashmi}. 
$D$-branes in plane wave background with nonconstant $NS-NS$ flux
can be obtained by applying $S$-duality on the above solutions.
We, however, will skip those details. In the next section we will 
analyze the supersymmetry of these solutions by solving type IIB 
Killing spinor equations explicitly.  
 
\sect{Supersymmetry Analysis}

The supersymmetry variation of dilatino and 
gravitino fields of type IIB supergravity in ten dimension, 
in string frame, is given by \cite{schwarz,fawad}:
\begin{eqnarray}
\delta \lambda_{\pm} &=& {1\over2}(\Gamma^{\mu}\partial_{\mu}\phi \mp
{1\over 12} \Gamma^{\mu \nu \rho}H_{\mu \nu \rho})\epsilon_{\pm} + {1\over
  2}e^{\phi}(\pm \Gamma^{M}F^{(1)}_{M} + {1\over 12} \Gamma^{\mu \nu
  \rho}F^{(3)}_{\mu \nu \rho})\epsilon_{\mp},\\
\label{dilatino}
\delta {\Psi^{\pm}_{\mu}} &=& \Big[\partial_{\mu} + {1\over 4}(w_{\mu
  \hat a \hat b} \mp {1\over 2} H_{\mu \hat{a}
  \hat{b}})\Gamma^{\hat{a}\hat{b}}\Big]\epsilon_{\pm} \cr
& \cr
&+& {1\over 8}e^{\phi}\Big[\mp \Gamma^{\mu}F^{(1)}_{\mu} - {1\over 3!}
\Gamma^{\mu \nu \rho}F^{(3)}_{\mu \nu \rho} \mp {1\over 2.5!}
\Gamma^{\mu \nu \rho \alpha \beta}F^{(5)}_{\mu \nu \rho \alpha
  \beta}\Big]\Gamma_{\mu}\epsilon_{\mp},
\label{gravitino}
\end{eqnarray}
where we have used $(\mu, \nu ,\rho)$ to describe the ten
dimensional space-time indices, and hat's represent the corresponding
tangent space indices. Solving the above two equations for the
$D5$-brane solution (\ref{D5}), we get several conditions on the
spinors. 

First the dilatino variation gives:
\bea
\Gamma^{\hat a} f_{5, \hat a} ~\epsilon_{\pm} + f^{-{1\over 4}}_5 
\Gamma^{\hat +\hat i\hat j} ~\p_{\hat i} b_{\hat j} (x_i) ~\epsilon_{\mp} 
+ {1\over 3!}~\Gamma^{\hat a\hat b \hat c}~\epsilon_{\hat a\hat b\hat c\hat d}
f_{5,\hat d} ~\epsilon_{\mp}= 0.
\label{dilaD5}
\eea 
On the other hand, the gravitino variation (\ref{gravitino}) gives
the following conditions on the spinors:
\bea
\delta \psi_+^{\pm} &\equiv & \p_{+}\epsilon_{\pm} + {1\over 4}f^{-{1\over
      4}}_5 \p _{\hat i} K(x_i) 
\Gamma^{\hat +\hat i} \epsilon_{\pm} - {1\over 8}f^{-{1\over 2}}_5
\Gamma^{\hat +\hat i\hat j}\p_{\hat i} b_{\hat j}(x_i) \Gamma^{\hat
  -}\epsilon_{\mp} = 0 
\label{G+}
\eea 
\bea
\delta \psi_-^{\pm} &\equiv & \p_{-}\epsilon_{\pm} = 0
\label{G-}
\eea
\bea
\delta \psi_i^{\pm} &\equiv & \p_{i}\epsilon_{\pm} - {1\over 8}f^{-{1\over
    2}}_5~\Gamma^{\hat +\hat j\hat
k}~\p_{\hat j}b_{\hat k}(x_i)\delta_{i\hat i} \Gamma^{\hat i} = 0  
\label{Gi}
\eea
\bea
\delta \psi_a^{\pm} &\equiv & \p_{a}\epsilon_{\pm} - {1\over
  8}{\p_{a}f_{5}\over f_5} \epsilon_{\pm} - {1\over 8}~\Gamma^{\hat +\hat
  i\hat j}~\p_{\hat i}b_{\hat j}(x_i)~\delta_{a\hat a}\Gamma^{\hat
  a}\epsilon_{\mp} = 0
\label{Ga}
\eea
In writing the above gravitino variation equations we have made use of 
the $D5$-brane supersymmetry condition:
\bea
\Gamma^{\hat a}\epsilon_{\pm} +{1\over 3!}~\epsilon_{\hat a\hat b\hat
  c\hat d}~\Gamma^{\hat b\hat c\hat d}\epsilon_{\mp} =0
\label{brsusy}
\eea
One notices that the supersymmetry condition (\ref{Gi}), for
nonconstant $F_3$: $\p_{\hat i}\p_{\hat j}
b_{\hat k} \ne 0$, can be satisfied only if
$\Gamma^{\hat +}\epsilon_{\pm} = 0$ \cite{russo}.

Using $\Gamma^{\hat +}\epsilon_{\pm} = 0$ and the brane supersymmetry
condition (\ref{brsusy}), the dilatino variation (\ref{dilaD5}) is
satisfied. Now, the supersymmetry condition (\ref{Ga}) is satisfied 
for the spinor $\epsilon_{\pm}$:~$\epsilon_{\pm} = \exp (-{1\over
  8}\ln f_5)\epsilon^0_{\pm}$, with $\epsilon^0_{\pm}$ being a function of
$x^+$ only. Since $\epsilon^0_{\pm}$ is independent of $x^i$ and $x^a$  
whereas $\partial_{\hat i} b_{\hat j}$ is a function of $x^i$ only, from
the gravitino variation (\ref{G+}), one gets the following 
conditions to have nontrivial solutions:
\bea 
\p_{\hat i} b_{j}(x_i) \Gamma^{\hat i\hat j}\epsilon^0_{\pm} = 0
\label{neces}
\eea
and 
\bea
\p_+ \epsilon^0_{\pm} = 0
\eea

For the particular case when $F_{+12} = F_{+34}$, the equation
(\ref{neces}) gives the following condition with constant spinor,
$\epsilon^0_{\pm}$:
\bea
\Gamma^{\hat 1\hat 2\hat3 \hat4}\epsilon^0_{\pm} = \epsilon^0_{\pm}.
\label{ppsusy}
\eea
Therefore the $D5$-brane solution (\ref{D5}) preserves 1/8
supersymmetry.

Now we analyze the supersymmetry of the system of two $D3$-branes as presented
in (\ref{d3-2}). The dilatino variation gives:
\bea
\Gamma^{\hat +\hat i\hat j}\p_{\hat i}b_{\hat j}(x_i)\epsilon_{\mp} = 0.
\label{dilad3}
\eea 
The gravitino variation gives the following conditions on the spinors 
to be solved:
\bea
\delta \psi_+^{\pm} &\equiv & \p_{+}\epsilon_{\pm} + {1\over 4}\p_{\hat i}
\Big( (1+X)^{-{1\over 4}} K(x_i)\Big)~\Gamma^{\hat + \hat i} -{1\over 8}
\p_{\hat i}b_{\hat j}(x_i)~\Gamma^{\hat +\hat i\hat j}\Gamma^{\hat -}
\epsilon_{\mp} \cr
& \cr
&{\mp}& {1\over 8} ~\Gamma^{\hat +\hat -\hat 6\hat 8\hat i}~\Bigg[
{{(1 + X_1 \sin^2 \alpha)^2 ~{\partial_i X_2} + \cos^2\alpha ~{\partial_i X_1}}
\over {(1 + X)^{3/2}(1 + X_1 \sin^2 \alpha)}}\Bigg]
\Gamma^{\hat +} \epsilon_{\mp} \cr
& \cr
&{\mp}& {1\over 8}  \Gamma^{\hat +\hat -\hat 5\hat 7\hat i}~\Bigg[
{1\over (1 + X)^{5/2}}(1 + X_1 \sin^2 \alpha) \cr
& \cr
&\times& ({{X_1}^2 \cos^2\alpha \sin^2\alpha }~{\partial_i X_2} + 
(1 + X_2)^2 \sin^2 \alpha ~{\partial_i X_1}) \cr 
& \cr
&-& {{({X_1}^2 \cos^2\alpha \sin^2\alpha)(~(1 + X_1 \sin^2 \alpha)^2 
~{\partial_i X_2} + \cos^2\alpha ~{\partial_i X_1})}\over (1 + X)^{5/2}
(1 + X_1 \sin^2 \alpha)} \cr
& \cr 
&+& {1\over{(1 + X)^{5/2}}}
\bigg(2{X_1}^2 \cos^2\alpha \sin^2\alpha ~{\partial_i X_2} + 
2{X_1}^3 \cos^2\alpha \sin^4\alpha ~{\partial_i X_1} \cr
& \cr
&-& 2 X_1(1 + X_2)\cos^2\alpha \sin^2\alpha ~{\partial_i X_1}\bigg)
\Bigg]\Gamma^{\hat +} \epsilon_{\mp}  \cr
& \cr 
&{\mp}& {1\over 8}~\bigg\{{{\Gamma^{\hat +\hat -\hat 5\hat 8\hat i}~
-~\Gamma^{\hat +\hat -\hat 6\hat 7\hat i}}
\over {(1 + X)^2( 1 + X_1 \sin^2\alpha)}}\bigg\} 
\Bigg[(- X_1\cos\alpha \sin \alpha ~{\partial_i X_2} \cr
& \cr
&+& (1 + X_2)\sin\alpha
\cos\alpha ~{\partial_i X_1} - {X_1}^2 \sin^3\alpha \cos\alpha
 ~{\partial_i X_1})( 1 + X_1 \sin^2\alpha) \cr
& \cr 
&+& X_1 \cos\alpha \sin\alpha
( 1 + X_1 \sin^2\alpha)^2~{\partial_i X_2} \cr
& \cr
&+& X_2\cos^3\alpha \sin\alpha ~{\partial_i X_1}\Bigg] 
\Gamma^{\hat +} \epsilon_{\mp} = 0
\label{d3+}
\eea
\bea
\delta \psi_-^{\pm} &\equiv & \p_{-}\epsilon_{\pm} = 0,
\eea
\bea
\delta \psi_a^{\pm} &\equiv & \p_{a}\epsilon_{\pm} - 
{1\over 8}(1+X)^{1\over 4}
\p_{\hat i}b_{\hat j}(x_i)~\Gamma^{\hat +\hat i\hat j}~\Gamma_{a}
\epsilon_{\mp} = 0,~~~~(a = 5,..,8),
\eea
\bea
\delta \psi_i^{\pm} &\equiv & \p_{i}\epsilon_{\pm} - {1\over 8}{{\p_{i}X}
\over{(1+X)}} \epsilon_{\pm}
- {1\over 8}(1+X)^{1\over 2}
\p_{\hat j}b_{\hat k}(x_i)~\Gamma^{\hat +\hat j\hat k}\delta_{i\hat i}~
\Gamma^{\hat i}\epsilon_{\mp} = 0.
\label{d3i}
\eea
In writing down the above supersymmetry variations, we have made use
of the following conditions\cite{rashmi}:
\bea
(\Gamma^{\hat 5\hat 8} - \Gamma^{\hat 6\hat
7})\epsilon_{\mp} = 0,~~~~(\Gamma^{\hat 5\hat 7} + \Gamma^{\hat 6\hat
8})\epsilon_{\mp} = 0, \label{rotation}
\eea
\bea
\Gamma^{\hat +\hat -\hat 6\hat 8}\epsilon_{\mp} = \epsilon_{\pm},~~~~
\Gamma^{\hat +\hat -\hat 5\hat 7}\epsilon_{\mp} = \epsilon_{\pm}.\label{brane}
\eea
To explain further, the conditions written in (\ref{rotation}) comes from
the rotation between the two $D3$-branes and those in (\ref{brane}) are the
$D3$-brane supersymmetry conditions. It is rather straightforward to conclude 
the conditions written in eqns. (\ref{rotation}) and (\ref{brane}) are
infact two independent conditions, thereby breaking $1/4$ supersymmetry.
As explained earlier, the equation (\ref{d3i}), for nonconstant
$\p_{\hat j}b_{\hat k}$, can be solved by the spinor $\epsilon_{\pm}$:~ 
$\epsilon_{\pm} = \exp(-{1\over 8}\ln(1+X))\epsilon^0_{\pm}$, with
$\epsilon^0_{\pm}$ being a function of $x^+$, only if: 
\bea
\Gamma^{\hat +}\epsilon_{\pm} = 0.
\label{nes}
\eea
Now putting the condition (\ref{nes}), the dilatino variation is satisfied.
All the gravitino variations are also satisfied leaving the following two 
equations to have nontrivial solutions.
\bea
\p_{\hat i}b_{\hat j}(x_i)\Gamma^{\hat i\hat j}\epsilon^0_{\mp} = 0.\label{pp}
\eea
and
\bea
\p_{+}\epsilon^0_{\pm} = 0,
\eea
Once again for the particular case $F_{+12} = F_{+34}$, eqn.(\ref{pp})
gives: $(1-\Gamma^{\hat 1\hat 2\hat
  3\hat4})\epsilon^0_{\mp} =0$ for constant spinor, $\epsilon^0_{\pm}$. 
Therefore the system of 
$D3$-branes (\ref{d3-2}) preserves $1/16$ supersymmetry \cite{rashmi}.

\sect{Summary and Discussion}

In this paper we have constructed various  
localized $D$-brane configurations in plane wave spacetime  
with nonconstant R-R 3-form flux. The supersymmetry of these
branes have been analyzed by solving type IIB Killing spinor
equations explicitly. The existence of other $Dp$-brane $(p<5)$ solutions 
in this plane wave spacetime puts restriction on the
localization of the branes and also on the behaviour of function $K(x_i)$
parameterizing the plane wave spacetime\cite{michelson,alis}. 
The ${\mathcal{H}}$-deformed 
$D$-branes can also be constructed following\cite{liu,ohta1,brecher}.
Though the nonextremal $D$-branes admit horizons and known as black branes,
this is not in general true in plane wave spacetime\cite{brecher1,liu,ohta1}. 
One could possibly look at the black brane solutions in this background
and discuss properties of their  horizon. 

The worldsheet construction 
of $D5$-brane and the corresponding nonlinear sigma model of the 
background considered in this paper can be found out by refering 
to the following Green-Schwarz action\cite{russo} written in lightcone gauge
and the $D5$-brane boundary condition:
\bea
L_B = \p_+ x_i \p_- x_i - {1\over 2} m^2 b^2_i + \p_+ y_a \p_- y_a, 
\eea 
\bea
L_f = i\theta_R \gamma^v \p_+ \theta_R + i\theta_L \gamma^v \p_-
\theta_L - {1\over 4} i m \p_i b_j(x_i)\theta_L \gamma^v \gamma^{i
  j}\theta_R,
\eea
\bea
m\equiv \alpha' p^u = \p_{\pm}u,
\eea
where $\theta_L$ and $\theta_R$ are the majorana-Weyl spinors in the
left and right moving sectors and $x^i,~(i=1,..4)$ and $y_a,~(a=5,..8)$
denote the worldvolume and transverse directions of the $D5$-brane 
respectively. The plane wave background with nonconstant R-R flux
can also be parametrized by holomorphic function on the 
worldsheet\cite{russo}. So it is useful to analyze the interacting 
Lagangian in the presence of these nonperturbative objects. 
The conditions of consistent $D$-brane which were obtained in
\cite{hikida} are expected to be different in the present case
because of the flat transverse
space. So an interesting exercise will be to obtain all the consistent
$D$-branes of \cite{hikida}. That would probably tell us about the 
integrability structure of the worldsheet theory in the presence of
branes, if it works out nicely, in a more general 
background. A systematic classification of all supersymmetric
$D$-branes from worldvolume point of view is also needed.
Finally, it would really be nice to find out the holographic dual
of these plane wave backgrounds in the presence of branes. 
We hope to come back to these issues in future.

\vspace{1cm}
\noindent
{\large \bf Acknowledgment:} The work of K.P. was supported in part by
I.N.F.N., by the E.C. RTN programs HPRN-CT-2000-00122 and
HPRN-CT-2000-00148, by the INTAS contract 99-1-590, by the MURST-COFIN
contract 2001-025492 and by the NATO contract PST.CLG.978785.



\newpage
\begin{thebibliography}{99}

\bibitem{penrose} R. Penrose, {\it ``Any space-time has a plane wave as a
  limit''},in Differential geometry and relativity, pp.271-275,
  Reidel, Dordrecht, (1976).

\bibitem{blau} M.Blau, J.Figuero-O'Farrill, C. Hull and 
G. Papadopoulos, JHEP {\bf 0201}, (2000) 047, hep-th/0110242,

M.Blau, J.Figuero-O'Farrill and G. Papadopoulos, Class. Quant. Grav.
{\bf 19, L87}(2000), hep-th/0201081,

M.Blau, J.Figuero-O'Farrill and G. Papadopoulos, 
Class.Quant.Grav. {\bf 19} (2002) 4753, hep-th/0202111.

\bibitem{malda} D. Berenstein, J. Maldacena, H. Nastase, JHEP {\bf
    0204} (2002) 013, hep-th/0202021.

\bibitem{maoz} J. Maldacena and L. Maoz, JHEP {\bf 0212} (2002) 046,
 hep-th/0207284.

\bibitem{russo} J. G. Russo and A. A. Tseytlin, JHEP {\bf 0209} (2002) 
035, hep-th/0208114.


\bibitem{hikida} Y. Hikida and S. Yamaguchi, JHEP {\bf 0301} (2003) 072,
hep-th/0210262.

\bibitem{kim} N. Kim, Phys. Rev. {\bf D67} (2003) 046005,
  hep-th/0212017.

\bibitem{bonelli}  G. Bonelli, JHEP {\bf 0301} (2003) 065,
  hep-th/0301089. 


\bibitem{pope} M. Cvetic, H. Lu, C.N. Pope, Nucl. Phys. {\bf B644} 
(2002) 65, hep-th/0203229.

M. Cvetic, H. Lu, C.N. Pope, K.S. Stelle, Nucl.Phys. {\bf B662} (2003) 89,
hep-th/0209193.

\bibitem{kumar} A. Kumar, R. R. Nayak, Sanjay, Phys. Lett. {\bf B541} 
(2002) 183, hep-th/0204025.

\bibitem{sken} K. Skenderis and M. Taylor, JHEP {\bf 0206} (2002) 025,
hep-th/0204054.

K. Skenderis, M. Taylor, Nucl. Phys. {\bf B665} (2003) 3, hep-th/0211011.

K. Skenderis, M. Taylor, JHEP {\bf 0307} (2003) 006, hep-th/0212184.

\bibitem{bain} P. Bain, P. Meessen and M. Zamaklar,  
Class. Quant. Grav. {\bf 20} (2003) 913, hep-th/0205106.

\bibitem{alis} M. Alishahiha and A. Kumar, Phys. Lett. {\bf B542}
 (2002) 130, hep-th/0205134.

\bibitem{zamak}  P. Bain, K. Peeters, M. Zamaklar, 
Phys.Rev. {\bf D67} (2003) 066001, hep-th/0208038.

\bibitem{kamal} A. Biswas. A. Kumar and K. L. Panigrahi,
Phys. Rev. {\bf D66} (2002) 126002, hep-th/0208042.

\bibitem{rashmi} R. R. Nayak, Phys.Rev. {\bf D67} (2003) 086006, 
hep-th/0210230.

\bibitem{alday} L. F. Alday, M. Cirafici, JHEP {\bf 0305} (2003) 006
      hep-th/0301253.

\bibitem{dabh} A. Dabholkar, S. Parvizi, Nucl. Phys.{\bf B641} 
(2002) 223, hep-th/0203231.

\bibitem{douglas} M. Berkooz, M. R. Douglas and R. G. Leigh, 
Nucl. Phys. {\bf B480} (1996) 265, hep-th/9606139.

\bibitem{gauntlett} J.P. Gauntlett, G.W. Gibbons, G. Papadopoulos
 and  P.K. Townsend, Nucl. Phys. {\bf B500} (1997) 133,
hep-th/9702202.

\bibitem{cvetic} K. Behrndt and M.Cvetic, Phys. Rev. {\bf D56} (1997) 
1188, hep-th/9702205.

\bibitem{myers} J.C. Breckenridge, G. Michaud and
 R.C. Myers, Phys. Rev.{\bf D56} (1997) 5172, hep-th/9703041.

\bibitem{hambli}  N.Hambli, Phys. Rev. {\bf D56} (1997) 2369, hep-th9703179.

\bibitem{leigh} V. Balasubramanian, F. Larsen and R.G. Leigh,
Phys. Rev.{\bf D57} (1998) 3509, hep-th/9704143.

\bibitem{townsend} P.K.Townsend,
Nucl. Phys. Proc. Suppl. {\bf 67} (1998) 88, hep-th/9708074.

\bibitem{ohta} N. Ohta and  P.K.Townsend,
Phys.Lett. {\bf B418} (1998) 77, hep-th/9710129.

\bibitem{alin}  A. Tirziu, P. Fendley, hep-th/0310074

\bibitem{kamal1} K. L. Panigrahi and Sanjay, Phys.Lett. B561 (2003) 284,
hep-th/0303182

\bibitem{liu} J. T. Liu, L. A. Pando Zayas, D. Vaman, hep-th/0301187.

\bibitem{ohta1} N. Ohta, K. L. Panigrahi and Sanjay, hep-th/0306186
(to appear in Nucl. Phys. B). 

\bibitem{myers1} J. C. Breckenridge, G. Michaud and  R. C. Myers,
Phys. Rev. {\bf D55} (1997) 6438, hep-th/9611174.

\bibitem{costa} M.S. Costa, G. Papadopoulos, 
Nucl. Phys. {\bf B510} (1998) 217, hep-th/9612204.

\bibitem{schwarz} J. H. Schwarz, Nucl. Phys.{\bf B226} (1983) 269. 

\bibitem{fawad} S.F. Hassan, Nucl.Phys. {\bf B568} (2000) 145, hep-th/9907152.

\bibitem{michelson} J. Michelson, Phys.Rev. {\bf D66} (2002) 066002,
hep-th/0203140.

\bibitem{brecher} D. Brecher, U. H. Danielsson, J. P. Gregory, M. E. Olsson,
hep-th/0309058. 

\bibitem{brecher1} D. Brecher, A. Chamblin, H.S. Reall, Nucl.Phys. {\bf B607} 
(2001) 155, hep-th/0012076.
\end{thebibliography}
\end{document}